\documentclass[twocolumn]{aastex631}

\usepackage{amsmath, xcolor, hyperref, makecell, enumitem}

\begin{document}

\title{Solar Wind Dependence on Source Distance from the Open-Closed Boundary}


\author[0000-0002-0154-8380]{Chloe P. Wilkins}
\affiliation{School of Science, University of Newcastle, University Drive, Callaghan, NSW 2308, Australia}

\author[0000-0002-1089-9270]{David I. Pontin}
\affiliation{School of Science, University of Newcastle, University Drive, Callaghan, NSW 2308, Australia}

\author[0000-0002-2728-4053]{Anthony R. Yeates}
\affiliation{Department of Mathematical Sciences, Durham University, Durham, DH1 3LE, UK}

\author[0000-0003-1692-1704]{Nicholeen M. Viall}
\affiliation{NASA Goddard Space Flight Center, Greenbelt, MD, USA}

\author[0000-0003-0176-4312]{Spiro K. Antiochos}
\affiliation{Department of Climate and Space Sciences and Engineering, University of Michigan, Ann Arbor, MI 48109, USA}

\begin{abstract}
The origin and variability of the slow solar wind remains an open question in solar physics, but is thought to be closely linked to dynamics at the Sun's open-closed magnetic flux boundary (OCB). Interchange magnetic reconnection at the OCB has been proposed as a mechanism for releasing closed-field plasma into the heliosphere, but observational evidence linking solar wind composition to OCB topology remains limited. 
We relate in situ solar wind measurements by Ulysses over a $\sim 10$-year period to the magnetic topology of their source regions using two coronal magnetic field models: a potential field source surface model and a magnetofrictional model.
We find a strong dependence of solar wind composition on the distance of the source magnetic flux from the OCB. Enhanced ion charge-state ratios, elemental abundances, and compositional variability are found to be concentrated within a supergranular-scale region (around 25~Mm) surrounding the OCB, consistent with the spatial scales of interchange magnetic reconnection. This variability decreases systematically with increasing distance from the boundary, with coronal hole wind exhibiting more uniform fast-wind signatures. We also find that the composition of solar wind emerging from regions close to the OCB is influenced by the strength of neighbouring closed magnetic fields, with stronger fields preferentially associated with slow-wind properties.
These results indicate that the composition of the slow wind is strongly governed by the magnetic topology of the OCB, providing compelling evidence that interchange reconnection plays a crucial role in slow solar wind release and structure.
\end{abstract}



\section{Introduction}\label{sec:Intro}

Beginning with some of the first in situ measurements of the solar wind \citep[e.g.,][]{Neugebauer_1962}, it has been known that the wind has many types, with two dominant forms: a “fast” wind with speeds of order 500 km/s and greater, and a “slow” solar wind with speeds of order 400 km/s or less. The fastest wind is generally considered to correspond to the wind predicted by \citet{Parker_1958} in his seminal work, because it is quasi-steady and appears to originate from the long-lived open magnetic field regions in the corona known as coronal holes \citep[e.g.,][]{Zirker_1977}. The origins of the remaining solar wind, including the canonical ``slow" wind, however, remain the subject of intense debate. Because slower wind is typically observed at low latitudes and in the vicinity of the heliospheric current sheet (HCS), proposed theories for its sources generally involve proximity to closed magnetic field regions in the corona---such as the closed field under the streamer belt---as well as active regions, the quiet Sun, and the boundaries of coronal holes. 

Two types of theories have been proposed for the release of plasma into the solar wind. In the first, the slow wind originates from quasi-steady open flux, much like the fast wind, but on flux that lies adjacent to closed-field regions. The associated magnetic flux tubes undergo super-radial expansion, with waves providing the energy required to accelerate the wind \citep[e.g.,][]{Kovalenko_1981, Wang_1991, Cranmer_2007}. In these models, the solar wind speed is largely determined by the flux-tube expansion factor, with larger expansion factors producing slower wind streams. 

The other type of theory is intrinsically dynamic, with versions of the idea referred to as the “streamer top” \citep{Suess_1996}, interchange magnetic reconnection \citep{Crooker_2004}, or dynamic S-Web \citep{Antiochos_2011} model. Here, magnetic reconnection between open and closed flux---i.e., interchange reconnection---leads to the release of denser, hotter plasma into the heliosphere, thereby forming the slow solar wind. The acceleration energy to the plasma may be provided by the reconnection process or by wave energy once the plasma is on open flux. At present, however, it remains unclear what fraction of the solar wind originates from each pathway. Nevertheless, because interchange reconnection must occur near the open-closed boundary (OCB), the magnetic topology of the OCB plays a critical role in all models for the slow wind. 

A promising approach for settling the debate of the origin of the slow wind is to establish the relationship between in situ solar wind composition and the magnetic topology at its source on the Sun. Compositional diagnostics, in particular ion charge-state ratios and the first ionisation potential (FIP) effect, provide powerful constraints on solar wind source regions and energisation mechanisms because they are set close to the Sun and remain largely unchanged during expansion into the heliosphere. The ion charge-state ratios $\textrm{C}^{6+}/\textrm{C}^{5+}$ and $\textrm{O}^{7+}/\textrm{O}^{6+}$ are inversely correlated with wind speed and commonly used to distinguish between types of wind streams \citep{Borovsky_2012}. These charge states are determined by the temperature, density and velocity of the plasma in the low to mid corona. Freeze-in distances calculated under the assumption of steady flow on open flux are found to occur below $\sim 2R_\odot$, where $R_\odot$ is the solar radius \citep{Landi_2012}. 

The FIP effect, by contrast, reflects elemental fractionation in the chromosphere, where low-FIP elements (e.g., Fe, Mg, Si) are separated from high-FIP elements (e.g., O, Ne) and preferentially transported into the corona \citep{Laming_2015, Pilleri_2015}. This results in coronal enhancements of low-FIP elements, which can be quantified through relative abundance ratios such as Fe/O. The slower wind exhibits a markedly stronger average FIP bias, and a stronger variability of the FIP bias than the faster solar wind. While the fast wind consists of near-photospheric abundances of low-FIP elements, the slow wind exhibits enhancements by factors of 3–4 \citep{Zurbuchen_1999}. This distinction is widely interpreted as evidence that the slow wind consists of plasma confined on closed magnetic field lines prior to release, suggesting that the magnetic topology of the OCB plays an important role in its structure. 

In this work, we determine the relationship between solar wind composition and coronal magnetic topology over $\sim 10$ years, using two global coronal magnetic field models together with in situ observations from Ulysses, whose unique high-latitude orbit sampled solar wind originating from deep within large polar coronal holes. In particular, we focus on the difference between solar wind associated with magnetic flux originating near the OCB---where interchange reconnection is expected to facilitate the release of closed-field plasma into the heliosphere---to solar wind associated with magnetic flux deep within a coronal hole. Section~\ref{sec:model_setup} introduces the inputs and simulation setup for the two coronal models considered in our study. Section~\ref{sec:model_observation} investigates the latitude-time evolution of the large-scale coronal magnetic field predicted by these models, and compares the results with in situ measurements by Ulysses. In Section~\ref{sec:mag_top}, we examine compositional signatures of the in situ solar wind and explore its dependence on the distance of the source magnetic flux from the OCB, and on the source magnetic field strength. Section~\ref{sec:conclusions} summarises our conclusions and discusses directions for future work. Together, this work provides new insight into the role of near-Sun magnetic topology---particularly in the vicinity of the OCB---in the release and compositional variability of the slow solar wind.

\section{Potential field extrapolations and time-dependent magnetofrictional simulations}\label{sec:model_setup}

To study the topology of the coronal magnetic field, we perform potential field source surface (PFSS) extrapolations and time-dependent magnetofrictional (TDMF) simulations spanning February 1993 to March 2003. Together, these provide complementary static and dynamic representations of the Sun's global magnetic structure. Each computation is performed on a grid of $60\times 180\times 360$ points in $(\ln r, \sin\lambda, \phi)$, where $r$, $\lambda$, and $\phi$ correspond to radius, latitude, and longitude, respectively.

\begin{figure}
    \epsscale{1.2}
    \plotone{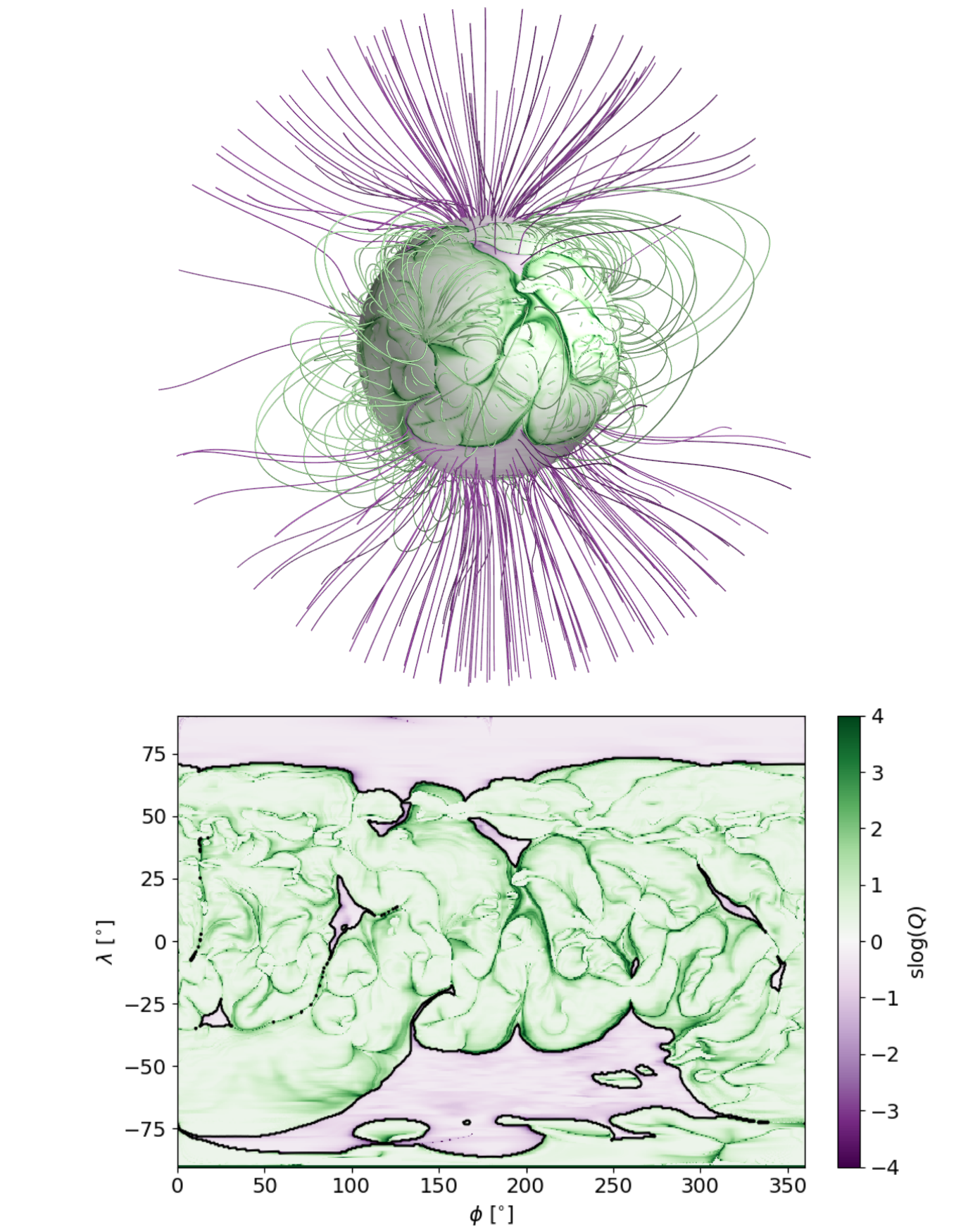}
    \caption{Top: a subset of field lines for the CR1863 PFSS extrapolation, with purple and green representing open and closed magnetic field connectivity, respectively. The signed logarithm of the squashing factor \citep[defined by][]{Titov_2002}, slog$(Q)$, is shown on the photosphere. This is defined to be
    positive (negative) when the field is closed (open). The slog$(Q)$ computations and field line tracing are performed using the Universal Fieldline Tracer \citep{UFiT_2024}. 
    Bottom: the corresponding latitude-longitude plot of slog$(Q)$ on the photosphere, with the OCB shown in black. 
    \label{fig:PFSS_3D_fieldlines}}
\end{figure}

The PFSS extrapolations are performed using the open-source package \texttt{pfsspy} \citep{pfsspy} with source-surface radius $R_{\rm ss}=2.5R_\odot$ \citep[following][]{Hoeksema_1983}. The input $B_r(R_\odot)$ is derived from National Solar Observatory (NSO) Kitt Peak Vacuum Telescope (KPVT) synoptic maps from the NASA/NSO Spectromagnetograph instrument \citep{NSO_KPVT}. For each extrapolation date (one per Carrington rotation, CR), $B_r(R_\odot)$ is constructed using the Durham Magnetofrictional Code (\texttt{DuMFriC}) pre-processing pipeline \citep{Yeates_2024}, which combines maps from neighbouring rotations to reduce edge effects. The maps are smoothed following \citet{Rice_2026}, then interpolated onto the computational grid and flux-balanced. PFSS solutions are computed from the resulting maps spanning CR1863 to CR1985. A subset of field lines from the PFSS extrapolation of the CR1863 map is shown in the top panel of \autoref{fig:PFSS_3D_fieldlines}. The bottom panel shows the latitude–longitude distribution of field line connectivity at the photosphere, with green and purple indicating open and closed field, respectively, and the OCB outlined in black.

\begin{figure*}
    \epsscale{1.15}
    \plotone{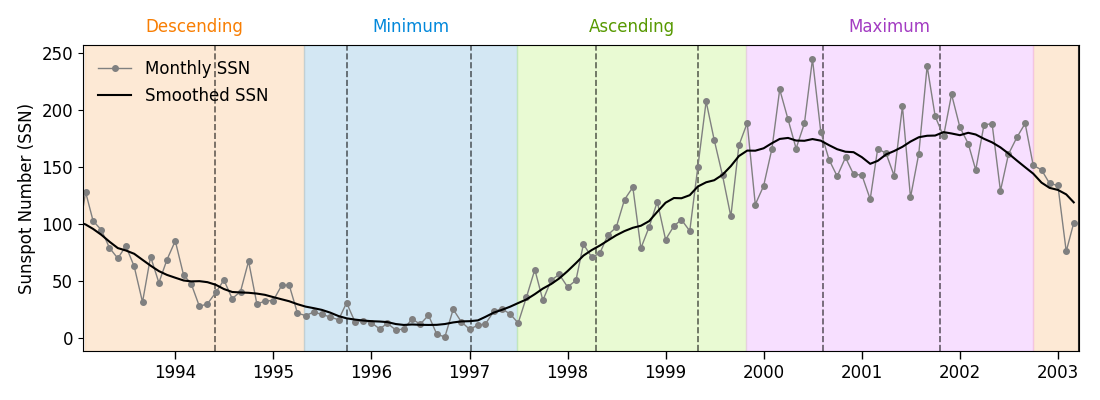}
    \caption{Monthly (grey) and smoothed (black) sunspot number (SSN) from 31 January 1993 to 19 March 2003, with the periods we define as \textit{descending}, \textit{minimum}, \textit{ascending}, and \textit{maximum} highlighted in orange, blue, green, and purple, respectively. Dashed black lines mark the transition times between the eight TDMF simulations. Sunspot number data provided by the U.S. Department of Commerce, NOAA Space Weather Prediction Center.
    \label{fig:sunspots}}
\end{figure*}

In contrast to the static PFSS extrapolations, the TDMF model evolves the coronal magnetic field using a driven photospheric boundary that incorporates surface flux transport (SFT) effects. Following \citet{Yeates_2024}, the non-ideal electric field term takes a hyperdiffusive form, and the plasma velocity is approximated using the magnetofrictional assumption combined with a radially outflowing wind profile. The boundary driving---which combines active region emergence, large-scale flows and supergranular diffusion---is applied through the horizontal surface electric field \citep[Eq.~4]{Yeates_2024}. Emerged active regions are identified using an automated procedure based on \citet{Yeates_2015} and \citet{Yeates_2020b}, and are inserted into the electric field following \citet[Eq.~5-6]{Yeates_2024} with $\tau=0.05$. 

The large-scale flows in the SFT scheme consist of a well-constrained differential rotation profile \citep{Snodgrass_Ulrich_1990} together with the meridional flow, 
\begin{align}\label{eq:meridional_flow}
    v_\theta(\theta) = D_u\cos\theta\sin^{p_0}\theta,
\end{align} 
where $D_u$ [km\,s$^{-1}$] and $p_0$ are constants. Unlike the differential rotation, optimal values for the supergranular diffusion---parameterised by the coefficient $\eta_0$ [km$^2$\,s$^{-1}$]---and meridional flow constants are less certain and may vary between solar cycles and with the specific set of emerged active regions. Our simulations adopt the parameters $\eta_0=466.8$, $p_0=2.33$ and $D_u=0.025$, which were optimised against observed butterfly diagrams over Solar Cycles~21-23 \citep{Whitbread_2018}.  

TDMF simulations are performed using \texttt{DuMFriC} on the same spatial grid as the PFSS extrapolations. Because the simulations incorporate only a subset of the full synoptic maps (i.e., the selected active regions), the magnetic field may gradually diverge from the underlying observations over time. Thus, rather than performing a single continuous 10-year run, we perform eight shorter simulations spanning approximately 18-month periods, saving the magnetic field solution approximately once per week. The transition times between each of the simulations are indicated by the vertical dashed lines in \autoref{fig:sunspots}. Each simulation is initialised from a PFSS extrapolation of a KPVT synoptic map, and requires a burn-in phase to reach a self-consistent magnetofrictional solution. Accordingly, we discard the first two months of each simulation from our analysis \citep[following][]{Hall_2025}. A detailed comparison of the PFSS and TDMF models, including their magnetic topology and open flux evolution over Solar Cycle 24, is given in \citet{Wilkins_2025}.

Over the $\sim 10$-year analysis period, we define four solar cycle phases to examine how the relationship between magnetic field topology and solar wind composition varies with global magnetic activity. These phases are shown alongside the sunspot number time series in \autoref{fig:sunspots}. The \textit{minimum} and \textit{maximum} phases correspond to periods where the smoothed sunspot number (black) is approximately $<30$ and $>150$, respectively. The \textit{ascending} phase spans the interval between, while the \textit{descending} phase covers the remaining periods leading into minimum and immediately following maximum. In the following section, we compare the model representations near solar minimum and maximum with in situ magnetic field observations from Ulysses. 

\section{Model-observation comparison of large-scale magnetic structure}\label{sec:model_observation}

To investigate the PFSS and TDMF representations of the Sun's large-scale magnetic field, we analyse the latitude–time evolution of the radial magnetic field and polarity inversion line at $2.5R_\odot$, and compare directly with in situ crossings of the HCS by Ulysses. Following \citet{Smith_1995}, crossings are identified from polarity reversals in the in situ radial magnetic field, $B_r$, obtained from NASA's Coordinated Data Analysis Web (CDAWeb) product \texttt{UY\_COHO1HR\_MERGED\_MAG\_PLASMA}\footnote{Median filtering is applied to $B_r$ prior to determining crossings. Information for the relevant CDAWeb products is available via the \href{https://cdaweb.gsfc.nasa.gov/misc/NotesU.html}{CDAWeb NotesU} page.} \citep{Br_Ulysses}. Periods of coronal mass ejection (CME) activity are identified using the \texttt{RICHARDSON} list \citep{Ul_Richardson} from the \texttt{ICMECAT} catalogue\footnote{\url{https://helioforecast.space/icmecat}} \citep{ICMECAT}, and crossings occurring during these intervals are excluded from the analysis. 

\begin{figure*}
    \epsscale{1.15}
    \plotone{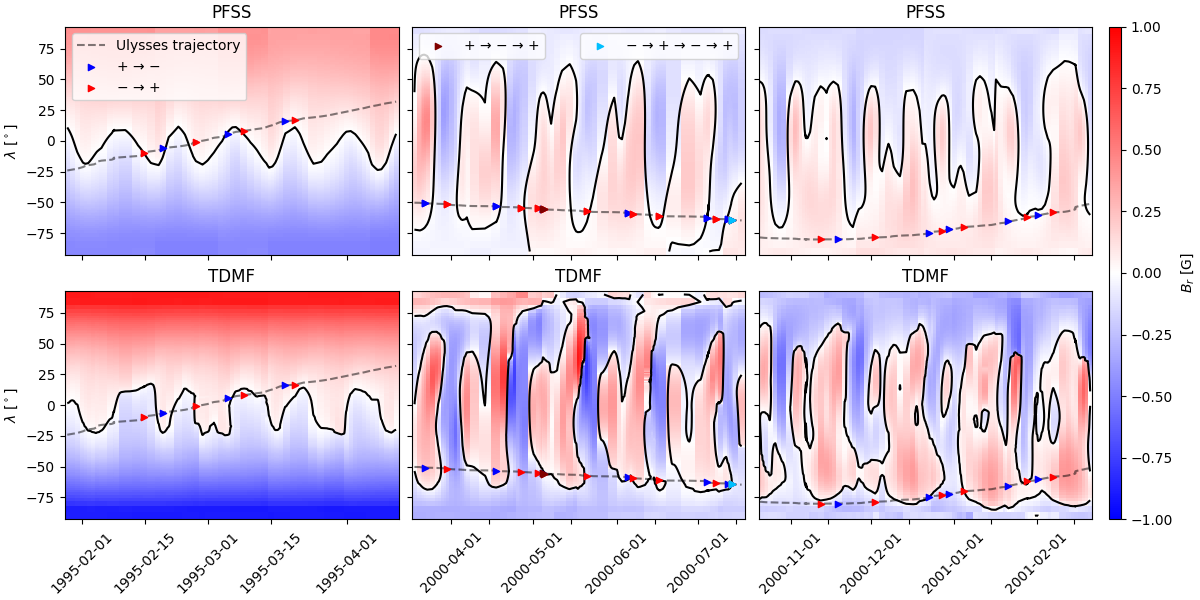}
    \caption{Latitude-time maps of the radial magnetic field and (black) polarity inversion line at $2.5R_\odot$ for the (top) PFSS model and (bottom) TDMF simulations. Ulysses' in situ position is back-mapped to $2.5R_\odot$; crossings from (red) negative to positive $B_r$ and (blue) positive to negative $B_r$ are overlaid along its trajectory (dashed grey line). Panels from left to right correspond to February-April 1995, April-July 2000, and November 2000-February 2001.
    \label{fig:HCS_lat_time}}
\end{figure*}

Since the observed solar magnetic field beyond $\sim 2.5R_\odot$ is open and effectively “frozen” into the outflowing plasma, each in situ measurement by Ulysses can be associated with a corresponding set of open magnetic field lines rooted at the Sun. Ulysses' daily-averaged heliographic position is obtained from the CDAWeb product \texttt{ULYSSES\_HELIO1DAY\_POSITION}. The daily-averaged velocity is determined from the radial component of the ion velocities in \texttt{UY\_M1\_BAI} \citep{UY_M1_BAI}, measured using Ulysses' Solar Wind Observations Over the Poles of the Sun (SWOOPS) instrument. For each day, we estimate the back-mapped time and source longitude of the associated solar wind parcel at $r=2.5R_\odot$ using a ballistic mapping method that assumes constant radial outflow and constant latitude for a given plasma parcel \citep{Neugebauer_1998, Parenti_2022}. The travel time is computed from the spacecraft position and ion velocity. The source longitude is then determined by accounting for solar rotation during the travel time, using the rotation rate $\Omega = 2.87\times10^{-6}$~rad\,s$^{-1}$ \citep{Parenti_2022}. Over the $\sim 1$–$5$~AU distance range of Ulysses, this is estimated to introduce a maximum longitudinal uncertainty of $<4^\circ$ \citep{Dakeyo_2024}, corresponding to only a few grid cells in the model magnetic fields. 

For each daily measurement, we determine the PFSS and TDMF model snapshots closest in time to the back-mapped position at $r=2.5R_\odot$. At the associated back-mapped longitude, the radial magnetic field at each latitude is extracted, yielding a single longitudinal strip, $B_r(\lambda)$. Over a specified time interval, these strips are concatenated to form a latitude-time map of the radial magnetic field at $r=2.5R_\odot$ corresponding to Ulysses' position. Using these maps, the evolution of the HCS is then approximated using the zero-contour of $B_r$.

Latitude-time maps of the HCS are shown for three periods in \autoref{fig:HCS_lat_time}: (i) February-April 1995, during Ulysses' first fast-latitude scan; (ii) April-July 2000, corresponding to the first sunspot maximum peak in \autoref{fig:sunspots}; and (iii) November 2000-February 2001, the middle of the maximum phase. For period (i), we identify seven in situ HCS crossings by Ulysses, consistent with those reported by \citet[Fig.~1]{Smith_1995} to within approximately 1-2 days. For periods (ii) and (iii), we identify 11 and 17 crossings, respectively (after excluding periods associated with CMEs). Ulysses' position at the resulting crossings is back-mapped to $r=2.5R_\odot$ using the ballistic method, with transitions from negative to positive $B_r$ marked by red arrows, and positive to negative transitions marked in blue. For period (ii), there are two instances where multiple crossings are detected within a short time window; these are distinguished in the plot by maroon and cyan markers.

Near solar minimum (see leftmost panels of \autoref{fig:HCS_lat_time}), the TDMF simulations produce a stronger polar magnetic field at $r=2.5R_\odot$ (which is driven largely by observational uncertainties near the poles) and a more structured HCS than the PFSS extrapolations, although the overall evolution is broadly consistent between the models. The HCS is largely equatorial and relatively flat, and the best agreement with the observed crossings is achieved by the PFSS model, consistent with the near-potential nature of the coronal field (at least on large scales) during this period.

During solar maximum (middle and right panels), larger discrepancies emerge between the PFSS and TDMF solutions---particularly at higher latitudes---as the global field becomes increasingly non-potential in the TDMF model. The discrepancies arise from a combination of differences between the underlying photospheric magnetograms at a given time, and the coronal field evolution (noting that the TDMF model builds and dissipates free energy in the corona over time through the surface driving). For period (ii), the PFSS model often shows good agreement with observations, capturing several of the polarity changes in the in situ magnetic field (e.g., the first four crossings). However, in other cases---such as the pair of crossings in early June 2000---the polarity changes are not reproduced. 

These differences become more pronounced in period (iii), where the PFSS model predicts a relatively simple global source-surface field with distinct north-south polarity regions. In contrast, TDMF simulations of this period capture the lingering influence of remnant polar fields prior to polarity reversal and better reproduce the observed high-latitude HCS structure. It is worth noting that the removal of CME periods is imperfect and, near solar maximum, the retained crossings are not always alternating in polarity. While this may indicate that some CME-related polarity reversals have been retained or that genuine HCS crossings have been inadvertently removed, it can also occur when Ulysses skirts the HCS and returns to the same side without fully traversing the current sheet. As such, one cannot expect a one-to-one correspondence between crossings identified in the data and those predicted by the models.

Overall, the PFSS model reproduces the latitude–time evolution of the HCS well during solar minimum, but often exhibits large discrepancies with observations near solar maximum. The TDMF simulations improve agreement in some instances, but still fail to capture certain observed crossings. Motivated by these findings, our subsequent analysis of coronal magnetic topology considers both the PFSS and TDMF models.

\section{Magnetic topology and solar wind composition from Ulysses}\label{sec:mag_top}

Building on the above model-observation comparison, we now investigate how the magnetic topology predicted by the PFSS and TDMF models relates to in situ solar wind properties measured by Ulysses. In particular, we examine compositional diagnostics---$\alpha$-particle velocities, ion charge-state ratios, and elemental abundances---to determine whether wind originating from magnetic flux rooted near the Sun's OCB differs systematically from wind originating deeper within coronal holes. 

\subsection{Observations of the in situ solar wind}\label{subsec:processing_SWICS}

To begin, we briefly describe the method used to relate in situ solar wind measurements to their source regions, and outline the data products used in our analysis. Between 31 January 1993 and 19 March 2003 (excluding periods of CME activity), daily-averaged observations of the solar wind by Ulysses are back-mapped to $r=2.5R_\odot$ using the ballistic method described in Section~\ref{sec:model_observation}. The corresponding magnetic footpoint (i.e., the location of the field line on the photosphere) is determined by tracing each open field line to $r=R_\odot$ using closest-in-time snapshots from the PFSS and TDMF models, and the Universal Fieldline Tracer \citep[\texttt{UFiT};][]{UFiT_2024}. 

To quantify the proximity of the source magnetic field to the OCB, we compute the great-circle distance between each open-field footpoint and the nearest closed-field footpoint. The nearest closed-field locations are obtained from photospheric maps of open-closed magnetic connectivity (e.g., see the bottom panel of \autoref{fig:PFSS_3D_fieldlines}) computed using \texttt{UFiT} on a $360\times 720$ grid uniformly spaced in latitude and longitude.

Over the $\sim 10$-year analysis period, in situ solar wind measurements from the Solar Wind Ion Composition Spectrometer (SWICS) onboard Ulysses are obtained from CDAWeb. Specifically, we consider \texttt{UY\_M1\_SCS} \citep{Geiss_Gloeckler_2025}, which provides 3.5-hour averages of:
\begin{itemize}
    \item $\alpha$ velocity (km\,s$^{-1}$), used as a proxy for the proton bulk speed;
    \item $\textrm{C}^{6+}/\textrm{C}^{5+}$;
    \item $\textrm{O}^{7+}/\textrm{O}^{6+}$;
    \item Fe/O, which compares a low-FIP (Fe) and high-FIP (O) element.
\end{itemize}
In addition, we consider three ratios representative of different FIP regimes (noting that these depend on both ionisation state and elemental abundance) from \texttt{UY\_M1\_SWI} \citep{Gloeckler_Geiss_2025}:
\begin{itemize}
    \item $\alpha/\textrm{O}^{6+}$ (helium; high FIP);
    \item $\textrm{Mg}^{10+}/\textrm{O}^{6+}$ (magnesium; low FIP);
    \item $\textrm{S}^{10+}/\textrm{O}^{6+}$ (sulfur; intermediate FIP).
\end{itemize}
Daily averages (centred on the spacecraft time) are then computed, yielding 3,400 valid days for \texttt{UY\_M1\_SWI} and 3,451 days for \texttt{UY\_M1\_SCS}. These products serve as compositional diagnostics of the solar wind, allowing us to distinguish between solar wind of different origins and examine how wind composition varies with the properties of its source magnetic field.

\subsection{Proximity of the source flux to the open-closed boundary}\label{subsec:dist_OCB}

\begin{figure}[h]
    \epsscale{1.15}
    \plotone{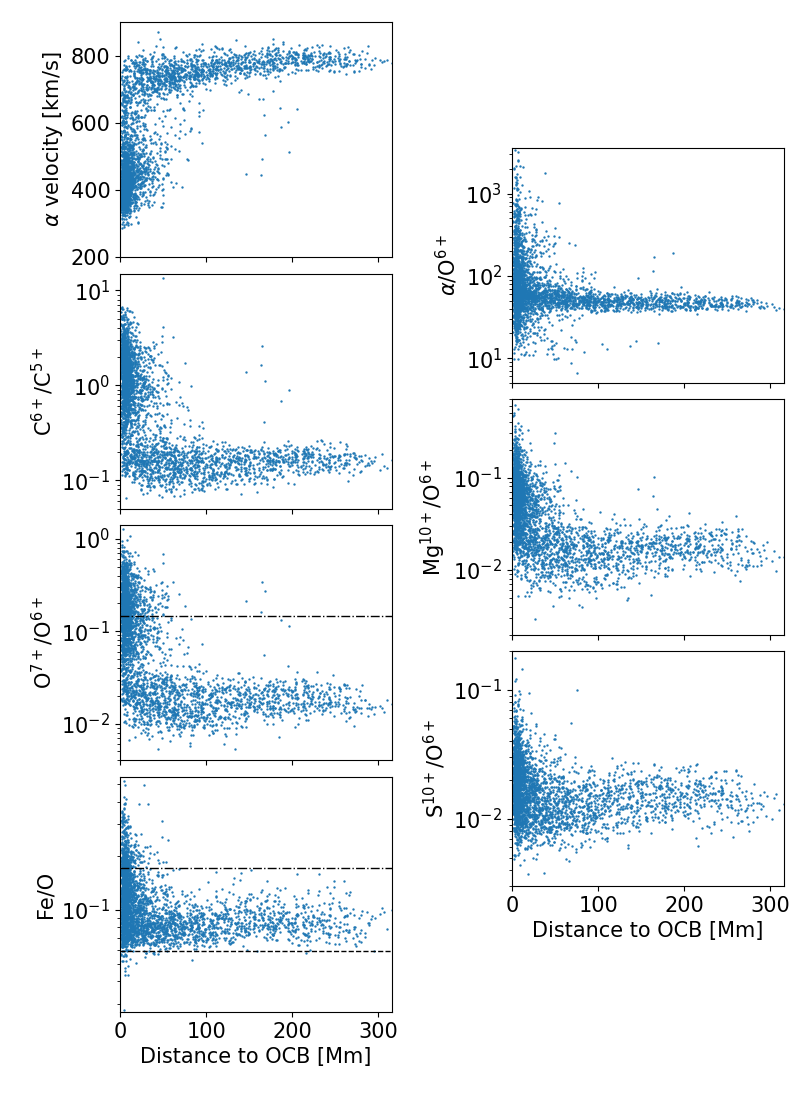}
    \caption{Distribution of SWICS diagnostics as a function of the distance of the source magnetic field from the OCB on the photosphere for the PFSS model. Coronal reference values (dash-dot lines) for $\textrm{O}^{7+}/\textrm{O}^{6+}$ and Fe/O, as well as the photospheric reference value (dashed line) for Fe/O are shown \citep{Zhao_2009, Schmelz_2012, Asplund_2021}.
    \label{fig:SWICS_dist_PFSS}}
\end{figure}

Motivated by the hypothesis that interchange reconnection contributes to the release of the solar wind, this section examines how in situ wind composition varies with the distance of the source magnetic flux from the OCB on the photosphere. For the PFSS model, distributions of the above SWICS diagnostics as a function of this distance are shown in \autoref{fig:SWICS_dist_PFSS}. Photospheric and coronal reference values for Fe/O are 0.059 \citep[dash-dot line;][]{Asplund_2021} and 0.17 \citep[dashed line;][]{Schmelz_2012}, respectively. The coronal reference value for $\textrm{O}^{7+}/\textrm{O}^{6+}$ is 0.145 (dash-dot line), with $\textrm{O}^{7+}/\textrm{O}^{6+}<0.145$ used by \citet{Zhao_2009} as a criterion for identifying solar wind originating from the centres of coronal holes. 

Similar plots constructed using the TDMF simulations yield broadly consistent trends, with the primary difference being a larger range of footpoint distances from the OCB. In the TDMF models, open-field footpoints typically lie farther from the OCB than in the corresponding PFSS extrapolations, reflecting the generally larger coronal holes---and greater unsigned open flux---in the TDMF solutions \citep{Wilkins_2025}. For brevity, we focus our initial analysis on the PFSS results.

\autoref{fig:SWICS_dist_PFSS} reveals that compositional signatures in the in situ wind vary with the proximity of their source magnetic flux to the OCB. The distributions reveal two broad populations:
\begin{enumerate}
    \item a \textit{variable} population close to the OCB, exhibiting a wide range of speeds and compositional signatures, encompassing the low $\alpha$ velocities, high charge states and enhanced Fe/O abundances characteristic of the canonical slow wind; and 
    \item a more \textit{steady} population that spans the full range of OCB distances (out to $\sim 300$~Mm), characterised by relatively uniform properties---wind speeds of $\sim 800$~km/s, low charge-state ratios, and small $\alpha/\textrm{O}^{6+}$ and Fe/O abundances.
\end{enumerate} 

A natural explanation for these trends is interchange reconnection. Close to the OCB, the broad range of compositional signatures may be explained by the dynamic mixing of closed- and open-field plasma. Consistent with this, wind originating from magnetic flux farther from the OCB---where such reconnection cannot occur---exhibits more steady compositional signatures characteristic of the classical fast solar wind. These results suggest that the transition from variable to steady wind compositions occurs over some finite distance from the OCB. An important question, therefore, is at what distance from the OCB this transition occurs.

\subsection{The transition from variable to steady wind}\label{subsec:transition}

\begin{figure*}
    \epsscale{1.15}
    \plotone{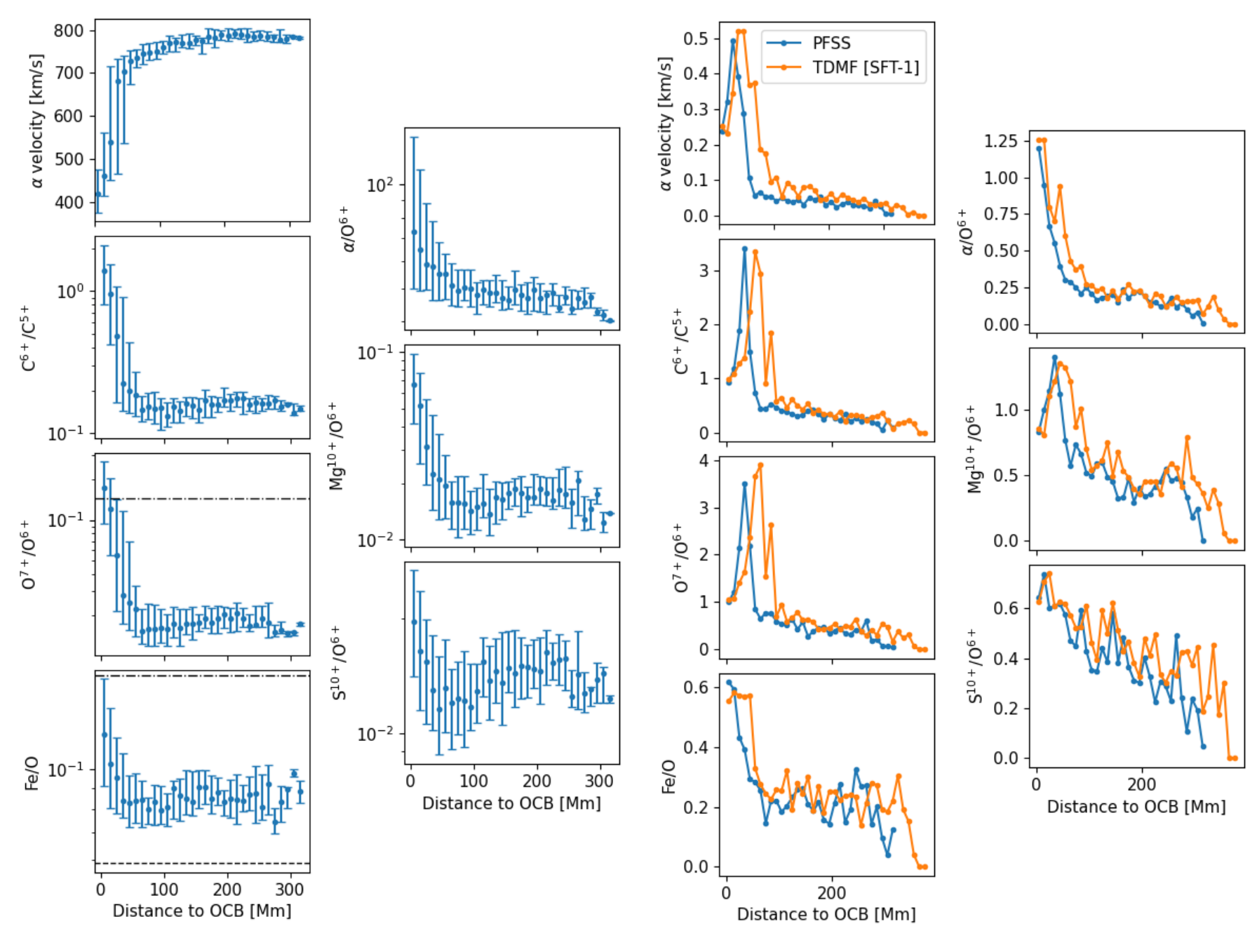}
    \caption{Distribution of SWICS diagnostics binned in 10~Mm intervals of distance from the OCB. Left: the median value in each bin, with error bars given by the first and third quartiles (for PFSS only). Right: the interquartile range normalised by the median value in each bin, for the PFSS (blue) and TDMF (orange) models. Reference values for coronal (dash-dot lines) and photospheric (dashed lines) abundances are shown as per \autoref{fig:SWICS_dist_PFSS}.
    \label{fig:dist_IQR}}
\end{figure*}

To address this question, we begin by examining the spread in the SWICS measurements of plasma composition as a function of distance from the OCB. For both the PFSS and TDMF models, we bin the data into 10~Mm intervals, and compute the median and interquartile ranges within each bin. The left panel of \autoref{fig:dist_IQR} shows the median values (for the PFSS model only), with error bars representing the first and third quartiles. The right panel shows the corresponding relative spread for both the PFSS and TDMF models, defined as the interquartile range normalised by the median.

Evidently, variability in the in situ wind composition is greatest for wind rooted near the OCB and generally decreases with distance from this boundary. The TDMF and PFSS models exhibit broadly consistent trends: the variability in the Fe/O, $\textrm{S}^{10+}/\textrm{O}^{6+}$ and $\alpha$ abundances is largest very close to the OCB. The relative spread in $\alpha$ velocities and the $\textrm{O}^{7+}/\textrm{O}^{6+}$, $\textrm{C}^{6+}/\textrm{C}^{5+}$, and $\textrm{Mg}^{10+}/\textrm{O}^{6+}$ charge-state ratios is also greater closer to the OCB, but peaks at slightly larger distances. For the PFSS model, this peak occurs at $\sim 40$~Mm, compared to 50-70~Mm for the TDMF model. 

The differences in relative spreads between the abundance ratios and the charge states may reflect the distinct chromospheric and coronal processes that determine these quantities, and their sensitivity to factors such as the frequency or height of reconnection. At present, however, the evolution of charge states under different reconnection scenarios and the response of FIP biases to reconnection-driven mixing between open- and closed-field plasma remains relatively unexplored. As a first step towards investigating these effects, we examine in Appendix~\ref{app:height} how the in situ charge states and FIP bias vary with the apex height of neighbouring closed field lines, but identify no clear systematic dependence within the uncertainties associated with our models.

\subsection{Clustering analysis of the variable wind populations}\label{subsec:clustering}

Complementing the above analysis of the relative spread in the SWICS distributions, we present a method for identifying the variable wind populations in \autoref{fig:SWICS_dist_PFSS} and estimating the characteristic distance from the OCB within which these populations are confined. We apply the \texttt{HDBSCAN} density-clustering algorithm \citep{scikit-learn} to each of the SWICS distributions, identifying clusters (with \texttt{min\_cluster\_size} = 15), and defining the variable population as the largest cluster in each case. 

\begin{figure}
    \epsscale{1.2}
    \plotone{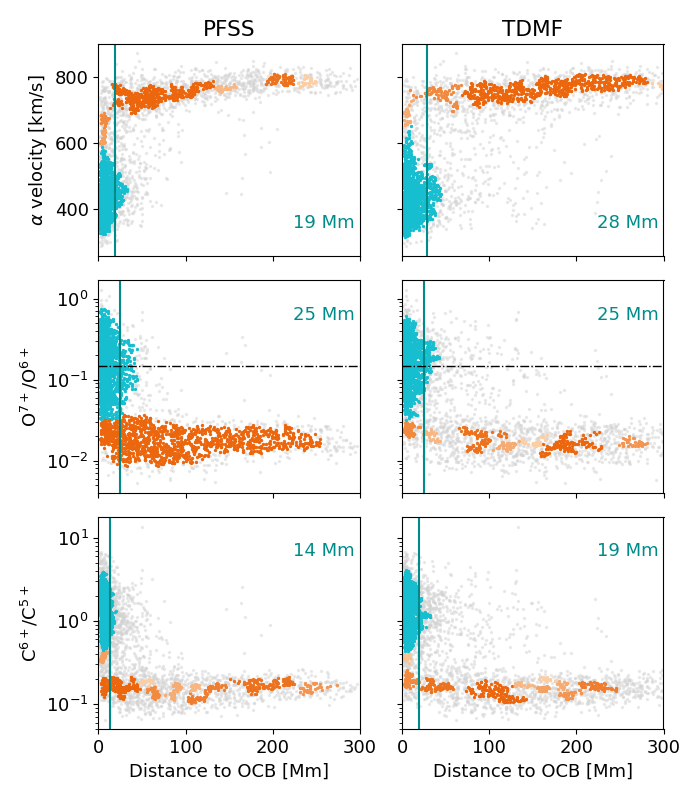}
    \caption{Clustering of (top) $\alpha$ velocities, (middle) $\textrm{O}^{7+}/\textrm{O}^{6+}$, and (bottom) $\textrm{C}^{6+}/\textrm{C}^{5+}$ as a function of distance from the OCB ($x \leq 300$~Mm). The left and right panels show results from the PFSS model and TDMF simulations, respectively. The cyan cluster identifies the variable wind population and the blue line marks the 95th percentile of the Gamma fit to this cluster. The remaining clusters are shown in orange (with colour intensity decreasing with cluster size) and noise (i.e., the set of points not identified as part of any cluster) is shown in grey. Reference values for coronal abundances (dash-dot line) are shown as per \autoref{fig:SWICS_dist_PFSS}.
    \label{fig:clustering}}
\end{figure}

The result of this procedure for the $\alpha$ velocities and two charge-state ratios is illustrated in \autoref{fig:clustering}, with the variable population shown in cyan and remaining clusters coloured orange. The variable cluster is broadly consistent across the two magnetic field models, with the TDMF simulations typically predicting a larger spatial extent from the OCB. In each case, the variable population contains 58--76\% of the total clustered data, while the steady population typically comprises smaller clusters and non-cluster points, described in the context of the clustering algorithm as noise (grey).

To determine a characteristic distance within which the bulk of the variable populations is confined, we construct frequency histograms of the variable clusters in 1~Mm bins, fit a Gamma distribution, and compute the 95th percentile (shown by the blue lines in \autoref{fig:clustering}) -- see Appendix \ref{app:clustering_Gamma} for details. The characteristic distance estimates for each of the SWICS diagnostics are summarised in \autoref{table:empirical_distance_estimates}. The results for $\alpha/\textrm{O}^{6+}$ and Fe/O (PFSS) are omitted as the variable population could not be reliably distinguished in these cases (see Appendix~\ref{app:clustering_Gamma}). 

\begin{deluxetable}{ccc}[h]
    \tablewidth{0pt}
    \caption{Empirical estimates of the characteristic distance (in Mm) from the OCB constraining 95\% of the variable solar wind populations across each magnetic field model.}
    \label{table:empirical_distance_estimates}
    \tablehead{
    \colhead{\makecell[tc]{Characteristic distance [Mm]}} & \colhead{\makecell[tc]{PFSS}} & \colhead{\makecell[tc]{TDMF}}}
    \startdata
    $\alpha$ velocity & 19 & 28 \\
     Fe/O & --- & 32 \\
     $\textrm{C}^{6+}/\textrm{C}^{5+}$ & 14 & 19 \\
     $\textrm{O}^{7+}/\textrm{O}^{6+}$ & 25 & 25 \\
     $\textrm{Mg}^{10+}/\textrm{O}^{6+}$ & 13 & 23 \\
     $\textrm{S}^{10+}/\textrm{O}^{6+}$ & 21 & 34
    \enddata
\end{deluxetable}

As expected given differences in coronal hole morphologies between the models, the characteristic distances are generally larger in the TDMF simulations than the PFSS model. The largest discrepancies occur for $\textrm{Mg}^{10+}/\textrm{O}^{6+}$ and $\textrm{S}^{10+}/\textrm{O}^{6+}$, with differences of up to $\sim 17$~Mm between the PFSS and TDMF results. Consistent with the relative spreads in \autoref{fig:dist_IQR}, the Fe/O abundances exhibit somewhat larger characteristic distances than the charge-state ratios, which may reflect the different heights at which these properties are set. 

Significantly, across all cases, the characteristic distance estimates suggest that the bulk of the variability occurs within approximately one supergranular diameter of the OCB \citep[$\sim 20$-40~Mm; e.g.,][]{Rieutord_2010, Schunker_2024}. This corresponds to the region within which interchange reconnection is thought to facilitate plasma exchange between open and closed flux \citep[e.g.,][]{Koukras_2023, Arge_2024}. We note that the characteristic distances should be interpreted as approximate indicators rather than sharp thresholds, as the inferred values exhibit some sensitivity to parameter choices. Differences associated with the adopted Gamma percentile are relatively modest, with the 85th and 95th percentiles changing the inferred distances by less than $\sim8$~Mm across the models. Variations in \texttt{min\_cluster\_size} produce larger changes; however, the inferred distances remain within approximately one supergranular diameter of the OCB (e.g., varying \texttt{min\_cluster\_size} between 15 and 30 yields maximum average distances across the charge states of 25~Mm and 40~Mm for the PFSS and TDMF models, respectively).

The suggestion that the bulk of the compositional variability occurs within approximately one supergranular diameter of the OCB is further supported by our work in \citet{Wilkins_2025}, where we showed that the fraction of open flux within 25~Mm of the OCB correlates closely with the observed fraction of slow wind in the heliosphere across Solar Cycle 24. Since compositional signatures of the slow wind (e.g., enhanced Fe/O abundances and charge-state ratios) are predominantly concentrated within the variable population, this in turn suggests that interchange reconnection likely plays an important role in both the release and variability of the slow wind.

Motivated by these findings, we focus our remaining analysis on the in situ composition of solar wind originating within 25~Mm of the OCB. 

\begin{figure*}
    \epsscale{1.15}
    \plotone{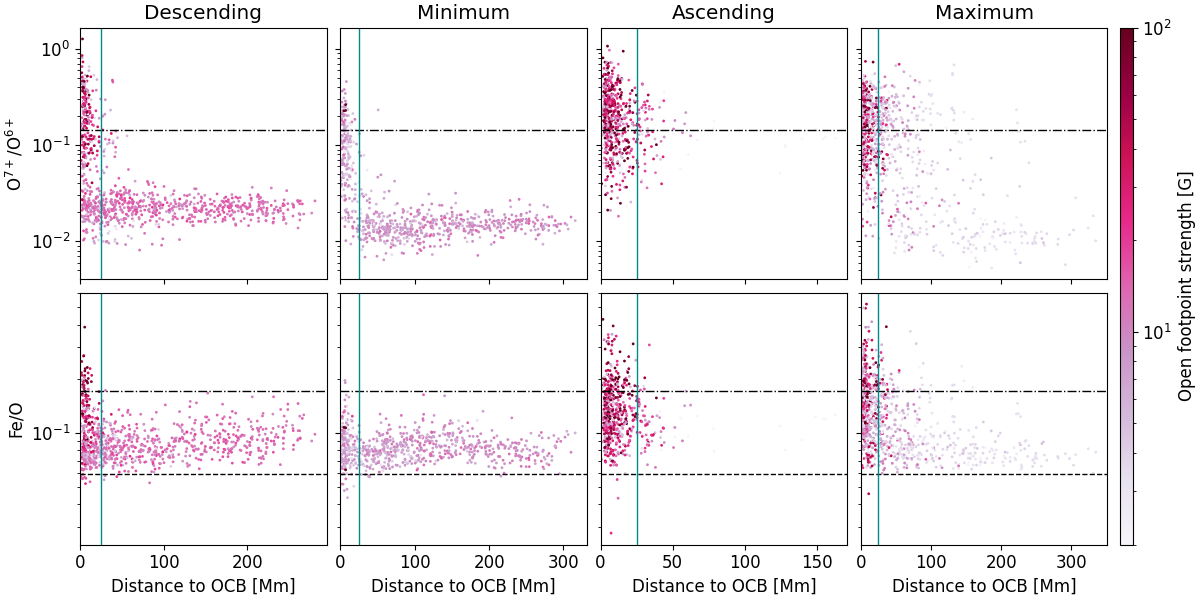}
    \caption{Distribution of (top) $\textrm{O}^{7+}/\textrm{O}^{6+}$ and (bottom) Fe/O in each solar cycle phase as a function of distance from the OCB, coloured by the photospheric footpoint strength of the corresponding back-mapped open field line. The blue line marks 25~Mm from the OCB. The results for the descending and minimum phases use the PFSS model, whilst the ascending and maximum phases use the TDMF model. Reference values for coronal (dash-dot lines) and photospheric (dashed lines) abundances are shown as per \autoref{fig:SWICS_dist_PFSS}.
    \label{fig:dist_strength_phases}}
\end{figure*}

\subsection{Wind originating from strong versus weak flux}\label{subsec:footpoint_strength}

\begin{figure*}
    \epsscale{1.15}
    \plotone{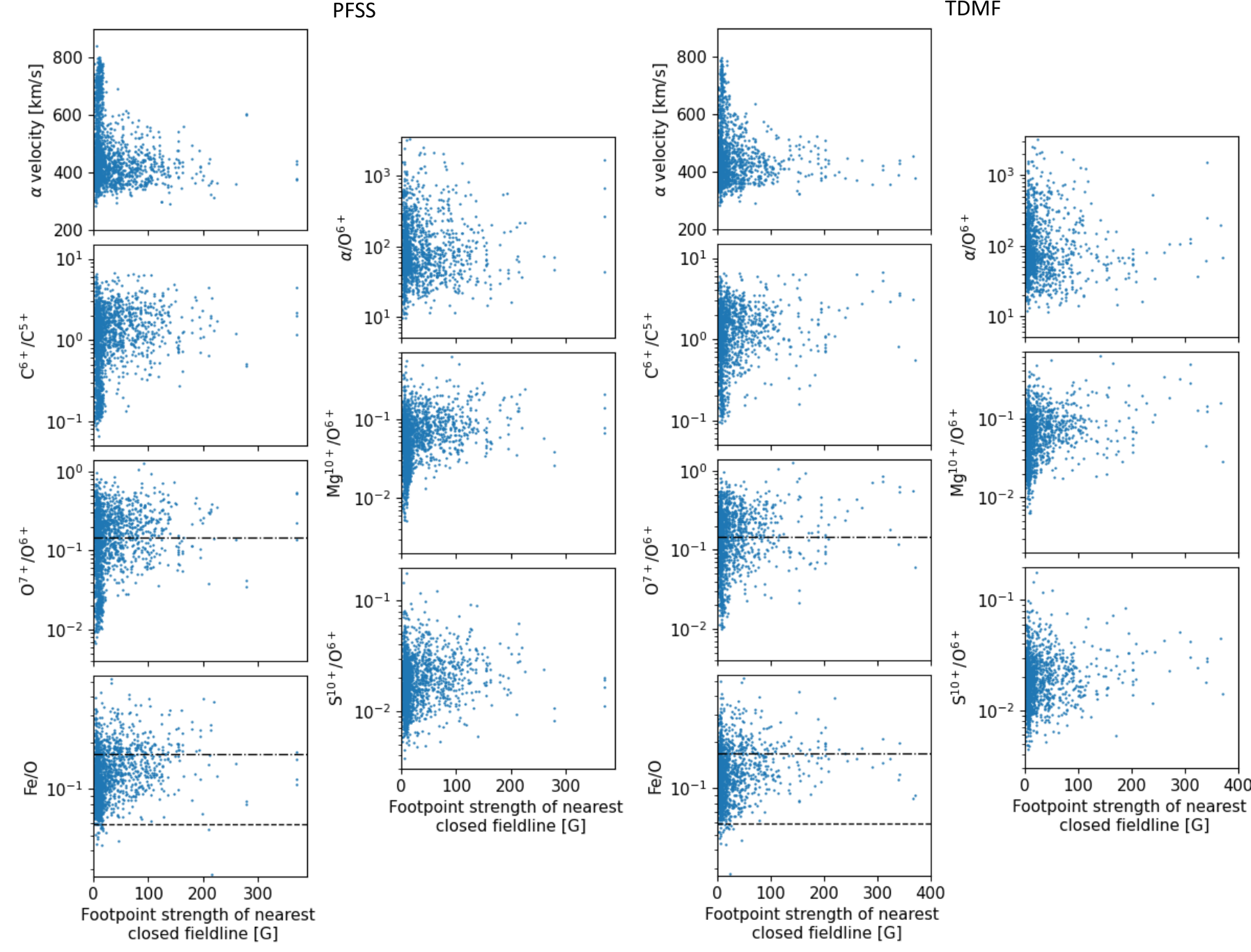}
    \caption{Distribution of SWICS diagnostics as a function of the photospheric footpoint strength of the nearest closed field line, restricted to wind that originates within 25~Mm of the OCB. The left and right panels show results for the PFSS model and TDMF simulations, respectively. For the TDMF simulations, the $x$-axis is restricted to 400~G (beyond which there are only three data points) to better illustrate trends. Reference values for coronal (dash-dot lines) and photospheric (dashed lines) abundances are shown as per \autoref{fig:SWICS_dist_PFSS}.
    \label{fig:SWICS_strength}}
\end{figure*}

Observations and models show that active region peripheries are associated with slower speed wind, and in situ measurements show that solar wind associated with active regions  exhibit the greatest value and range of FIP biases, hottest charge state ratios, and highest relative alpha abundances \citep[e.g.,][]{Kasper_2007, brooks2015}. Furthermore, the distinctive composition of active region plasma, relative to that of the quiet Sun, is thought to be consistent with an increased occurrence of interchange reconnection \citep{Macneil_2019}. However, the distribution of magnetic field strengths near the OCB---where interchange reconnection occurs---remains relatively unexplored. In this section, we investigate how the in situ composition of the solar wind changes with the strength of the source magnetic field within 25~Mm of the OCB.

For each back-mapped open field line, the photospheric magnetic field strength at the corresponding footpoints is determined. Since the prevalence and strength of active regions varies over the solar cycle---typically peaking near solar maximum and diminishing toward minimum---we begin by examining the results within each cycle phase. \autoref{fig:dist_strength_phases} shows the distribution of $\textrm{O}^{7+}/\textrm{O}^{6+}$ charge states and Fe/O abundances as a function of the distance of the corresponding open field line from the OCB in each phase, coloured by the photospheric footpoint strengths. The results in the descending and minimum phases use the PFSS model, while the ascending and maximum phases use the TDMF simulations, motivated by the results described in Section \ref{sec:model_observation}.  
Most of the stronger-field regions are confined within 25~Mm of the OCB (indicated by the blue line), consistent with observations of coronal hole boundaries adjacent to active regions. These regions are ubiquitous during the ascending phase due to Ulysses' position in the equatorial plane (helio-latitudes between $-40^\circ$ and $8^\circ$). During the descending and maximum phases, they are closely associated with solar wind exhibiting enhanced $\textrm{O}^{7+}/\textrm{O}^{6+}$ ratios, and Fe/O abundances near the coronal reference value (black dash-dot line), although photospheric Fe/O abundances are also occasionally observed.

We then restrict our analysis to field lines that map to within 25~Mm of the OCB. Because active regions are primarily associated with closed magnetic loops, we characterise the local magnetic environment using the photospheric footpoint strength of the nearest closed field line (though using the open-field strength yields similar results due to the proximity to the OCB). 

\autoref{fig:SWICS_strength} shows the SWICS diagnostics as a function of the nearest closed footpoint strength for wind whose source magnetic field lies within 25~Mm of the OCB. Weaker closed magnetic field ($<30$~G) corresponds to a broad range of in situ compositions associated with both fast and slow wind (although this may also reflect increased uncertainty in the field line mapping in weak-field regions). Stronger ($>100$~G) nearby closed fields are typically associated with lower $\alpha$ velocities, enhanced charge-state ratios and coronal abundances of Fe/O, characteristic of the slow solar wind. This behaviour is consistent with the slow wind being preferentially released from coronal hole boundaries near strong active-region loops, where closed-field plasma can be transferred onto adjacent open flux via interchange reconnection. Overall, these results indicate that the composition of solar wind emerging from regions close to the OCB is influenced by the strength of neighbouring closed magnetic fields, with stronger fields preferentially associated with slow-wind properties.

\section{Conclusions}\label{sec:conclusions}

In this work, we combine global coronal magnetic field models with observations by Ulysses to relate compositional solar wind diagnostics to their source magnetic topology. The unique Ulysses orbit, which reached high latitudes and sampled deep within polar coronal holes, enables us to investigate solar wind originating both near and far from the OCB, and to examine the role of interchange reconnection in determining the compositional properties of the in situ plasma. 

We find a strong dependence of the in situ wind composition on the distance of the source magnetic flux from the OCB for both the PFSS and TDMF magnetic field models. Solar wind originating closest to the OCB exhibits the broadest range of speeds, charge states, and elemental abundances, encompassing both canonical fast and slow wind signatures. Solar wind originating from deep within the polar coronal holes exhibits comparatively uniform fast-wind compositions. The compositional variability in the in situ wind generally decreases with increasing distance from the OCB. Statistical clustering over the $\sim 10$-year analysis period indicates that the bulk of the variability occurs within approximately one supergranular diameter of the OCB, consistent with the spatial scales over which interchange reconnection is expected to facilitate plasma exchange between open and closed magnetic flux. Ultimately, this work reveals a strong link between in situ solar wind composition and the proximity of its source magnetic field to the OCB, highlighting the important role of OCB dynamics in shaping the observed structure and variability of the solar wind. 

Within a 25~Mm band of the OCB, the composition of the solar wind also depends on the strength of the magnetic flux at the corresponding footpoints. Open field rooted adjacent to strong closed magnetic field regions is associated with enhanced in situ charge states and elemental abundances, consistent with previous observations linking active-region coronal loops to slow solar wind release. In contrast, wind associated with weaker surrounding flux exhibits a broader range of compositional properties characteristic of both the fast and slow wind. While this may reflect intrinsic differences in solar wind associated with strong versus weak flux, these results may also be influenced by increased uncertainties in the back-mapping procedure in weaker field regions and near the OCB. 

We also examine whether in situ charge states and FIP bias vary with the apex heights of neighbouring closed field lines. Within the uncertainties of the present models, we find no clear systematic dependence.

The dependence of solar wind composition on field strength warrants further investigation, because it provides a promising method for distinguishing between wave acceleration on continuously open magnetic fields versus interchange theories discussed in Section~\ref{sec:Intro}. Through detailed studies of the plasma composition within the closed field of active regions adjacent to coronal holes---such as those performed by \citet{Wallace_2025}---it may become possible to quantitatively relate the composition of in situ plasma to that of its source region at the Sun. Such a result would provide strong support for models in which interchange reconnection plays a significant role in the release of the slow solar wind.

The comparison of global coronal magnetic field models with in situ observations highlights several important avenues for future work. Longitudinal uncertainties associated with ballistic back-mapping may contribute to discrepancies between the observed and modelled HCS, and may also influence the inferred photospheric source regions of the in situ solar wind. Quantifying the impact of these uncertainties, together with those arising from the coronal field models themselves, will therefore be important for future analyses. Further progress will also require improved magnetic field observations and better constraints on the TDMF model parameters. Improved polar flux measurements from missions such as Solar Orbiter, together with recent work using solar eclipse observations to optimise the imposed solar wind outflow profile in the TDMF model \citep{Rice_2026}, provide promising pathways for improving future simulations. Finally, while recent studies have explored differences in reconnection dynamics between streamer and pseudostreamer topologies \citep[e.g.,][]{Aslanyan_2022, Dey_2025}, comparatively little attention has been given to how these magnetic environments influence solar wind composition. Future studies combining global magnetic field models with squashing factor maps may provide insight into how solar wind composition varies between distinct coronal structures and with the height and topology of the nearest closed magnetic field.
\\

\section{Acknowledgments}

CW was supported by an Australian Government Research Training Program (RTP) Scholarship. DP gratefully acknowledges support through an Australian Research Council Discovery Project (DP210100709). 
SKA acknowledges support for this research from the NASA LWS and NSF SHINE Programs.
CW \& DP acknowledge the Awabakal people, the traditional custodians of the unceded land on which their research was undertaken at the University of Newcastle. 
ARY was supported by the Science and Technology Facilities Council (UKRI1216). NMV was supported by the competed Heliophysics Internal Scientist Funding Model.

\bibliography{references}{}
\bibliographystyle{aasjournal}


\appendix

\section{The density-clustering and Gamma fitting procedure}\label{app:clustering_Gamma}

This appendix provides additional details of the method used to determine the characteristic distances in Section~\ref{subsec:clustering}, which quantify the distance from the OCB within which significant variations in solar wind speed and composition are observed. We present the frequency histograms and fitted Gamma distributions for the variable cluster in the $\alpha$ velocity distribution, and highlight two cases that are excluded from the main-text analysis.

\begin{figure}[h]
    \epsscale{0.9}
    \plotone{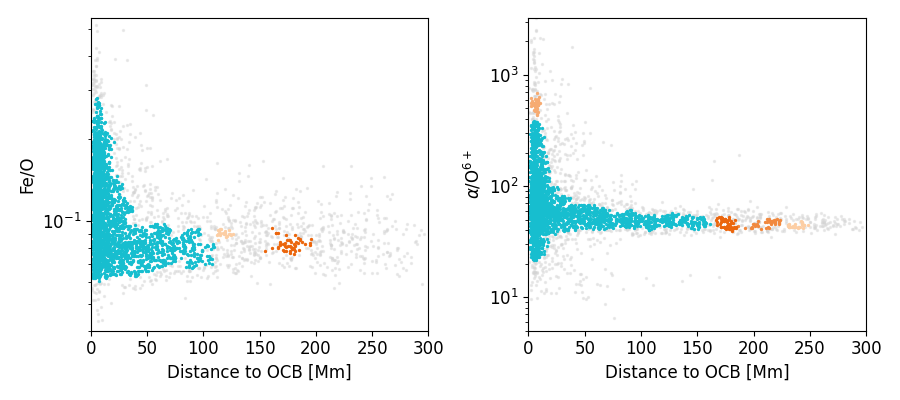}
    \caption{Results of the \texttt{HDBSCAN} clustering for the (left) Fe/O and (right) $\alpha/\textrm{O}^{6+}$ distributions as a function of distance from the OCB for the PFSS model (restricted to $x\leq 300$~Mm).}
    \label{fig:poor_clustering}
\end{figure}

As described in Section~\ref{subsec:clustering}, the variable population of the SWICS distributions as a function of distance from the OCB is determined by applying the \texttt{HDBSCAN} clustering algorithm with \texttt{min\_cluster\_size} = 15. For most datasets, this procedure yields a well-defined variable population. However, there are two notable exceptions: the $\alpha/\textrm{O}^{6+}$ distributions and the Fe/O distribution in the PFSS model. As shown in \autoref{fig:poor_clustering}, the identification of a distinct variable population is less clear for these quantities, particularly in the $\alpha/\textrm{O}^{6+}$ case. The variable cluster contains a substantial fraction of points that appear to overlap with the steady wind population. This may reflect the greater complexity of the processes governing elemental abundances and $\alpha$-particle composition, which may not naturally separate into the same two broad populations identified by the charge-state ratios. This is consistent with the suggestion by \citet{Roberts_2020} that a larger number of clusters (beyond the simple variable/steady dichotomy) may be required to fully describe solar wind compositional structure.

\begin{figure}[h]
    \epsscale{1.15}
    \plotone{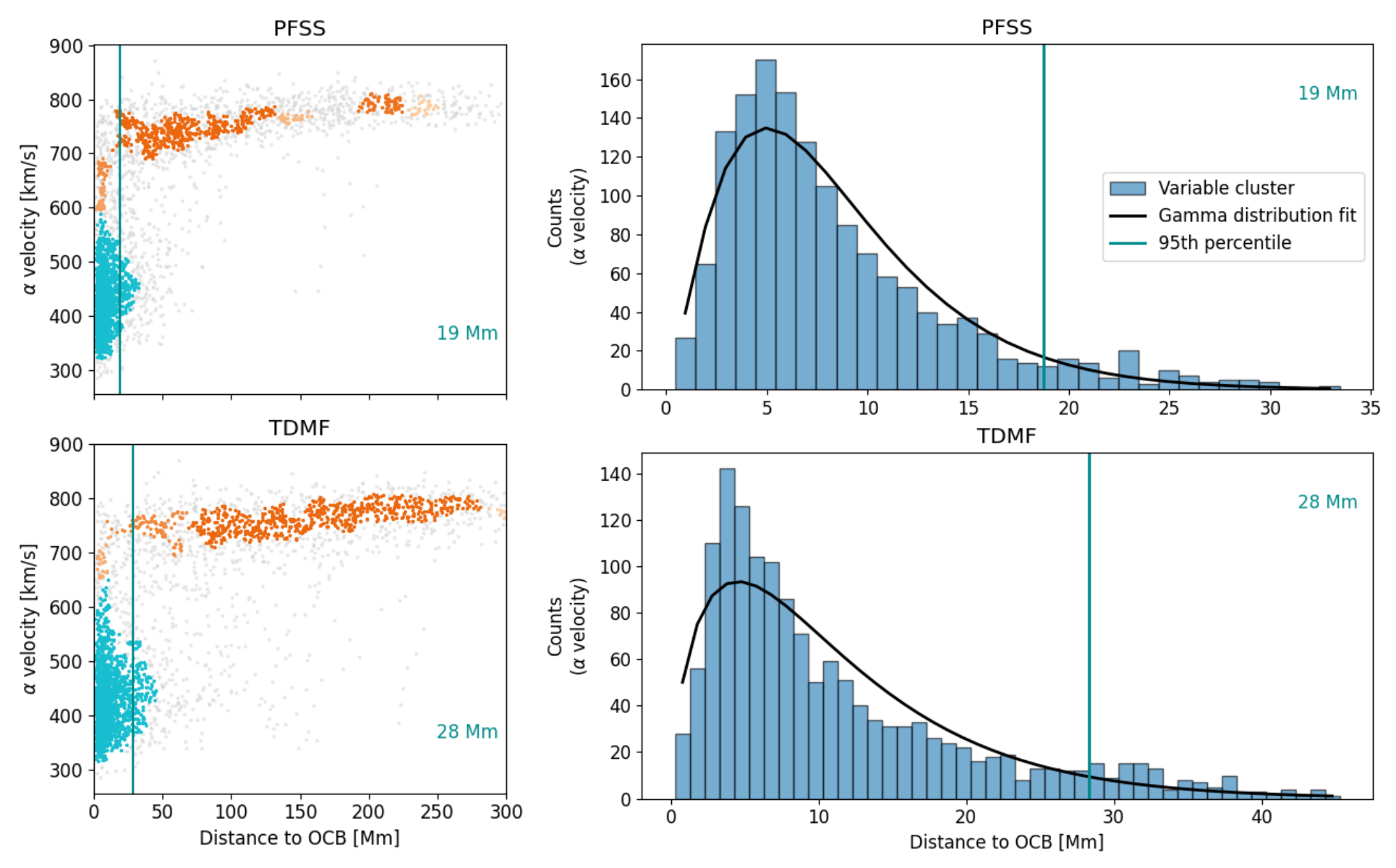}
    \caption{Left: clustering of $\alpha$ velocities as a function of distance from the OCB (restricted to $x\leq 300$~Mm). Right: frequency histograms of the variable (cyan) cluster in 1~Mm bins, with the 95th percentiles (blue) of the fitted Gamma distributions (black) indicated. The top and bottom panels correspond to the PFSS and TDMF models, respectively.}
    \label{fig:vel_clustering_and_gamma}
\end{figure}

Excluding the results for the $\alpha/\textrm{O}^{6+}$ and Fe/O (PFSS), the variable clusters for the remaining compositional diagnostics are then analysed. Frequency histograms are constructed in 1~Mm bins of the distance of the source magnetic flux from the OCB. The resulting histograms for the $\alpha$ velocity distribution are shown in the right panel of \autoref{fig:vel_clustering_and_gamma}. In each case, the histograms are right-skewed and are well described by a Gamma distribution (black curve), determined using \texttt{scipy.stat.gamma}. The 95th percentile of the fitted distribution is then used as an estimate of the characteristic distance from the OCB within which the bulk of the variable population is confined.

\section{The height of the nearest closed field line}\label{app:height}


Given the relationship between loop heights and electron temperatures and densities---which in turn determine the altitude at which charge states are frozen in---the size of coronal loops undergoing interchange reconnection may contribute to determining the charge-state composition of the slow wind \citep[e.g.,][]{Viall_2020}. To investigate this, we examine how in situ charge states and FIP bias varies with the apex height of the nearest closed field line for wind parcels mapping to within 25~Mm of the OCB.

\begin{figure}
    \epsscale{0.6}
    \plotone{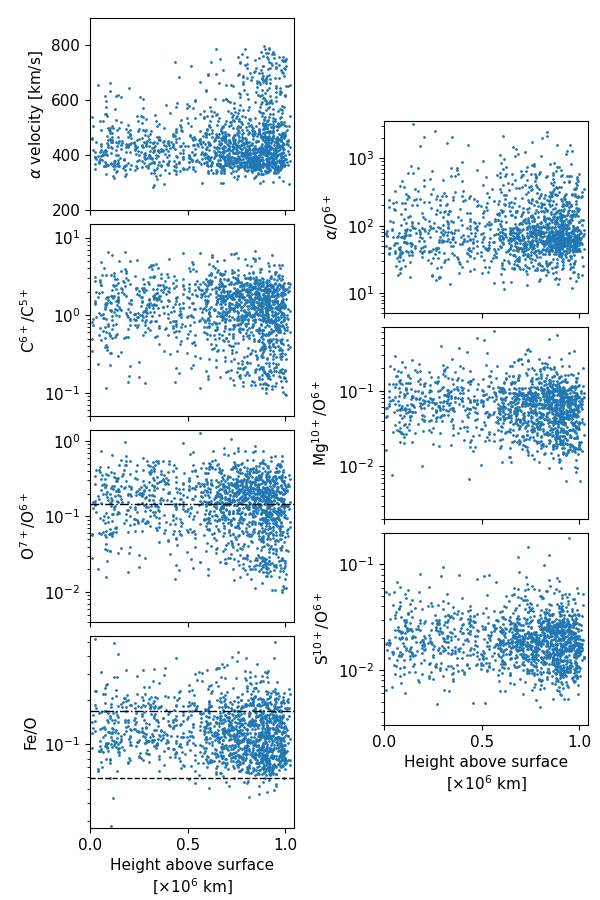}
    \caption{Distribution of SWICS diagnostics as a function of the apex height of the nearest closed field line above the photosphere for wind parcels that map to within 25~Mm of the OCB for the TDMF simulations. Reference values for coronal (dash-dot lines) and photospheric (dashed lines) abundances are shown as per \autoref{fig:SWICS_dist_PFSS}.
    \label{fig:Height_TDMF}}
\end{figure}

Starting from the photospheric footpoint of the nearest closed region, we trace the corresponding field line using \texttt{UFiT} and determine its apex height above the photosphere (in km). \autoref{fig:Height_TDMF} shows the resulting SWICS distributions as a function of apex height for the TDMF simulations. Distributions from the PFSS model yield similar results, though with a higher frequency of short closed field lines ($<0.25\times10^6$~km) due to the presence of smaller-scale polarity structures (which are progressively smoothed in the TDMF photospheric field by SFT evolution).

Low $\alpha$ velocities and enhanced ratios of $\textrm{O}^{7+}/\textrm{O}^{6+}$, $\textrm{C}^{6+}/\textrm{C}^{5+}$ and $\textrm{Mg}^{10+}/\textrm{O}^{6+}$ occur across the full range of apex heights, suggesting that slow wind signatures are associated with both short and long closed field lines. In contrast, Fe/O, $\textrm{S}^{10+}/\textrm{O}^{6+}$, and $\alpha/\textrm{O}^{6+}$ are relatively uniformly distributed across the neighbouring loop heights, showing no clear dependence on apex height. Fast solar wind signatures---characterised by lower $\textrm{O}^{7+}/\textrm{O}^{6+}$ and $\textrm{C}^{6+}/\textrm{C}^{5+}$, and higher $\alpha$ velocities---are associated with a narrower range of neighbouring loop heights and appear more frequently near longer field lines ($>0.75\times 10^6$~km). 

These trends in the in situ wind composition are less robust and less clearly defined than those discussed in Section~\ref{sec:mag_top}, namely the proximity of the field line to the OCB and the strength of the source magnetic field. The lack of clear trends here may reflect uncertainties in the back-mapping procedure and in the field line tracing in weak-field regions. Additionally, this may reflect limitations of the steady-state model assumptions, particularly the fixed freeze-in heights of the charge states and the prescribed magnetic field topology, when applied to an inherently time-dependent process. As such, we conclude that, while no distinct relationship between charge states and the apex heights of neighbouring closed field lines is observed here, further work is required to properly investigate this relationship.

\end{document}